\documentclass[aip, amsmath,amssymb,reprint]{revtex4-2}

\usepackage{graphicx}
\usepackage{dcolumn}
\usepackage{bm}

\usepackage[utf8]{inputenc}
\usepackage[T1]{fontenc}
\usepackage{mathptmx}
\usepackage{etoolbox}
\usepackage{makecell}
\usepackage{amsmath}
\usepackage{upgreek}
\usepackage{xr}

\makeatletter
\def\@email#1#2{%
 \endgroup
 \patchcmd{\titleblock@produce}
  {\frontmatter@RRAPformat}
  {\frontmatter@RRAPformat{\produce@RRAP{*#1\href{mailto:#2}{#2}}}\frontmatter@RRAPformat}
  {}{}
}%
\makeatother

\draft 

\begin{document}

\title{Ferroelectricity of Wurtzite Al$_{1-x}$Hf$_{x}$N Heterovalent Alloys} 

\author{Nate S.P. Bernstein}
 \affiliation{Department of Metallurgical and Materials Engineering, Colorado School of Mines, Golden, Colorado 80401, United States}
 \email{nbernstein@mines.edu}
 
\author{Daniel Drury}
 \affiliation{U.S. Army Combat Capabilities Development Command - Army Research Laboratory, Adelphi, Maryland 20783, United States}
 
\author{Cheng-Wei Lee}
 \affiliation{Department of Metallurgical and Materials Engineering, Colorado School of Mines, Golden, Colorado 80401, United States}

\author{Tatau Shimada}
 \affiliation{TAIYO YUDEN CO., LTD., 2-7-19, Kyobashi, Chuo-ku, Tokyo, 104-0031, Japan}
 
\author{Yuki Sakai}
 \affiliation{TAIYO YUDEN CO., LTD., 2-7-19, Kyobashi, Chuo-ku, Tokyo, 104-0031, Japan}

\author{Oliver Rehm}
 \affiliation{Department of Physics, Universit\"{a}t Konstanz, D-78457 Konstanz, Germany}

\author{Lutz Baumgarten}
 \affiliation{Forschungszentrum J\"{u}lich GmbH, Peter Grünberg Institut (PGI-6), D-52425 J\"{u}lich, Germany}

\author{Martina M\"{u}ller}
 \affiliation{Department of Physics, Universit\"{a}t Konstanz, D-78457 Konstanz, Germany}

\author{Prashun Gorai}
 \affiliation{Department of Chemical and Biological Engineering, Rensselaer Polytechnic Institute, Troy, New York 12180, United States}
 \affiliation{Department of Metallurgical and Materials Engineering, Colorado School of Mines, Golden, Colorado 80401, United States}
 
\author{Yoshiki Iwazaki}
 \affiliation{TAIYO YUDEN CO., LTD., 2-7-19, Kyobashi, Chuo-ku, Tokyo, 104-0031, Japan}
 
\author{Glen R. Fox}
 \affiliation{Fox Materials Consulting, LLC, Colorado Springs, Colorado 80908, United States}
 
\author{Keisuke Yazawa}
 \affiliation{Department of Metallurgical and Materials Engineering, Colorado School of Mines, Golden, Colorado 80401, United States}
 \affiliation{Materials Science Center, National Renewable Energy Laboratory, Golden, Colorado 80401, United States}
 
\author{Brendan Hanrahan}
 \affiliation{U.S. Army Combat Capabilities Development Command - Army Research Laboratory, Adelphi, Maryland 20783, United States}
 
\author{Geoff L. Brennecka}
 \affiliation{Department of Metallurgical and Materials Engineering, Colorado School of Mines, Golden, Colorado 80401, United States}
 \email{gbrennec@mines.edu}


\begin{abstract}
Thin films of aluminum hafnium nitride (Al$_{1-x}$Hf$_{x}$N) were synthesized via reactive magnetron sputtering for Hf contents up to $x$ = 0.13. X-ray diffraction showed a single $c$-axis oriented wurtzite phase for all films. Hard X-ray photoelectron spectroscopy demonstrated homogeneous Al:Hf distribution through the thin films and confirmed their insulating character. A collection of complementary tests showed unambiguous polarization inversion, and thus ferroelectricity in multiple samples. Current density vs.\ electric field hysteresis measurements showed distinct ferroelectric switching current peaks, the piezoelectric coefficient d$_{33,f,meas}$ measured using a double beam laser interferometer (DBLI) showed a reversal in sign with similar magnitude, and anisotropic wet etching confirmed field-induced polarization inversion. This demonstrates the possibility of using tetravalent--and not just trivalent--alloying elements to enable ferroelectricity in AlN-based thin films, highlighting the compositional flexibility of ferroelectricity in wurtzites and greatly expanding the chemistries that can be considered for future devices.
\end{abstract}

\maketitle 

Ternary wurtzite-type aluminum nitride based thin films (Al$_{1-x}$\textit{M}$_{x}$N) have gained recent attention directed at applications utilizing their ferroelectric (FE) behavior. These applications include, among others, high-operating temperature, nonvolatile, random-access memory (HOT-NVM)\cite{drury_high-temperature_2022} and ferroelectric high electron mobility transistors (Fe-HEMTs).\cite{casamento_ferrohemts_2022, casamento_chapter_2023} To date, the following Al$_{1-x}$\textit{M}$_{x}$N ferroelectric thin films have been reported: Al$_{1-x}$Sc$_{x}$N \cite{fichtner_alscn_2019}, Al$_{1-x}$B$_{x}$N \cite{hayden_ferroelectricity_2021}, Al$_{1-x}$Y$_{x}$N \cite{wang_ferroelectric_2023}, and Al$_{1-x}$Gd$_{x}$N.\cite{lee_prediction_2024} Note that the \textit{M}-element in all cases is nominally trivalent, a logical approach to replace the Al$^{3+}$ cation. 

The conventional expectation is that percent-level additions of a non-trivalent \textit{M}-element could introduce sufficiently high free carrier concentrations to lead to metallic conduction. In this work, we demonstrate that sputtered thin films of Al$_{1-x}$Hf$_{x}$N are electrical insulators and can be ferroelectric up to at least $x$ = 0.13. To date, this is the first Al$_{1-x}$\textit{M}$_{x}$N thin film showing ferroelectricity with heterovalent \textit{M}-element alloying. Importantly, this work shows that researchers working on HOT-NVM and FE-HEMTs are not limited to trivalent cation replacements, thus increasing compositional options for AlN-based ferroelectric films and driving deeper studies of charge balancing defect compensation in III-N alloys.

A charge-balancing approach has also dominated the more extensive alloying efforts across the piezoelectric thin film community, including both direct substitution of Al$^{3+}$ by other trivalent species and stoichiometric combinations of multiple alloying species.\cite{thomas_effect_2022} Such multivalent alloy studies started with computational work from Iwazaki\cite{iwazaki_highly_2015} and Tholander\cite{tholander_large_2015}, and Akiyama and coworkers have consistently led the community's experimental efforts, including key studies on Al$_{1-x}$(Mg,Nb)$_{x}$N \cite{uehara_giant_2017}, Al$_{1-x}$Mg$_{x/2}$Ti$_{x/2}$N \cite{anggraini_mg_2019}, Al$_{1-x}$(Mg,Ta)$_{x}$N \cite{anggraini_enhancement_2022}, and Al$_{1-x}$Mg$_{x/2}$Hf$_{x/2}$N.\cite{nguyen_development_2024} It is worth noting that Uehara's Al$_{1-x}$(Mg,Nb)$_{x}$N study reported the greatest piezoelectric response in films with Mg/Nb ratios that would not correspond to an effective average valence of 3+ for pure Mg$^{2+}$ and Nb$^{5+}$, but they did see evidence for multivalency in the Nb species.\cite{uehara_giant_2017} 

Studies on intentionally heterovalent alloys such as Al$_{1-x}$Mg$_{x}$N\cite{anggraini_effect_2018}, Al$_{1-x}$Si$_{x}$N\cite{anggraini_polarity_2020}, and even Al(O,N)\cite{akiyama_influence_2008,islam_improved_2024} suggest some degree of control over growth polarity. However, the focus of these reports has primarily been on piezoelectric properties, with only the report by Islam \textit{et al.} discussing polarization reversal and leakage current.\cite{islam_improved_2024} Reports of Al$_{1-x}$Sc$_{x}$N remaining electrically insulating\cite{akiyama_influence_2008} and even ferroelectric with an approximate oxygen content of 4 at.\% \cite{islam_improved_2024, drury_toward_2023} suggest value in exploring other donor dopants. DFT calculations of the piezoelectric response also motivate study of Al$_{1-x}$Hf$_{x}$N,\cite{startt_unlocking_2023} and heterovalent AlN-based alloys more broadly. 

Thin films of Al$_{1-x}$Hf$_{x}$N were synthesized using reactive magnetron sputtering techniques similar to previous studies \cite{yazawa_reduced_2021,drury_high-temperature_2022, drury_understanding_2021}. Primary samples were deposited on (001) 4H-SiC substrates with sputtered continuous molybdenum bottom and top electrodes (Mo/Al$_{1-x}$Hf$_{x}$N/Mo/SiC); this stack was chosen because of its relevance to high temperature electronics. The top electrodes were all circular and varied in diameter from 50 $\mu$m to 125 $\mu$m. Top electrode patterning was done with wet etching and a photolithographic liftoff process. Three sputter targets were used: Mo (99.95 \% Kurt J Lesker Co.), Al (99.999 \%, Kurt J Lesker Co.), and Hf (99.9 \% Stanford Advanced Materials). Targets were 10.16 cm in diameter. The Hf target purity was 99.5 \% when including Zr as a contaminant (i.e., Zr<0.5 \%). Two samples of Al$_{1-x}$Hf$_{x}$N were grown by varying the Hf target RF power while maintaining a constant Al target pulsed DC power to control the Al/Hf cation ratio (see Table~\ref{tab:table1}). One sample of AlN was grown in the same series to compare these films to previous work. The nitride layer of all three thin film stacks was sputtered with the following conditions: 2 mTorr of Ar/N$_2$ (40/40 sccm flow), a substrate heater set point of 400°C, and a deposition time of 100 min. The Hf content listed in Table~\ref{tab:table1} was measured using Rutherford Backscattering Spectrometry (RBS) made by National Electrostatics Corp. Elements other than Hf, Al, Mo, Zr, and Si were all below the detection threshold of the RBS instrument. Complementary samples for beamline measurements were deposited on AlN-seeded sputtered continuous molybdenum bottom electrodes on Si wafers. Additional preparation and synthesis details are provided in the Supporting Material.

\begin{table}
\caption{\label{tab:table1}Deposition conditions and film properties.}
\begin{ruledtabular}
\begin{tabular}{cc|cc}
\makecell{Hf Power Density\\(W/cm\textsuperscript{2})} & \makecell{Total Dep.\ Rate\\(nm/min)} & \makecell{Hf Cation Content\\(at. \%)} & \makecell{Film Thickness\\(nm)}\\
\hline
0.0& 1.8& 0& 180\\
11.8 & 2.0 & 6 & 200\\
17.7 & 2.2 & 13 & 220\\
\end{tabular}
\end{ruledtabular}
\end{table}

\begin{figure*}
\centering
\includegraphics[scale=0.35]{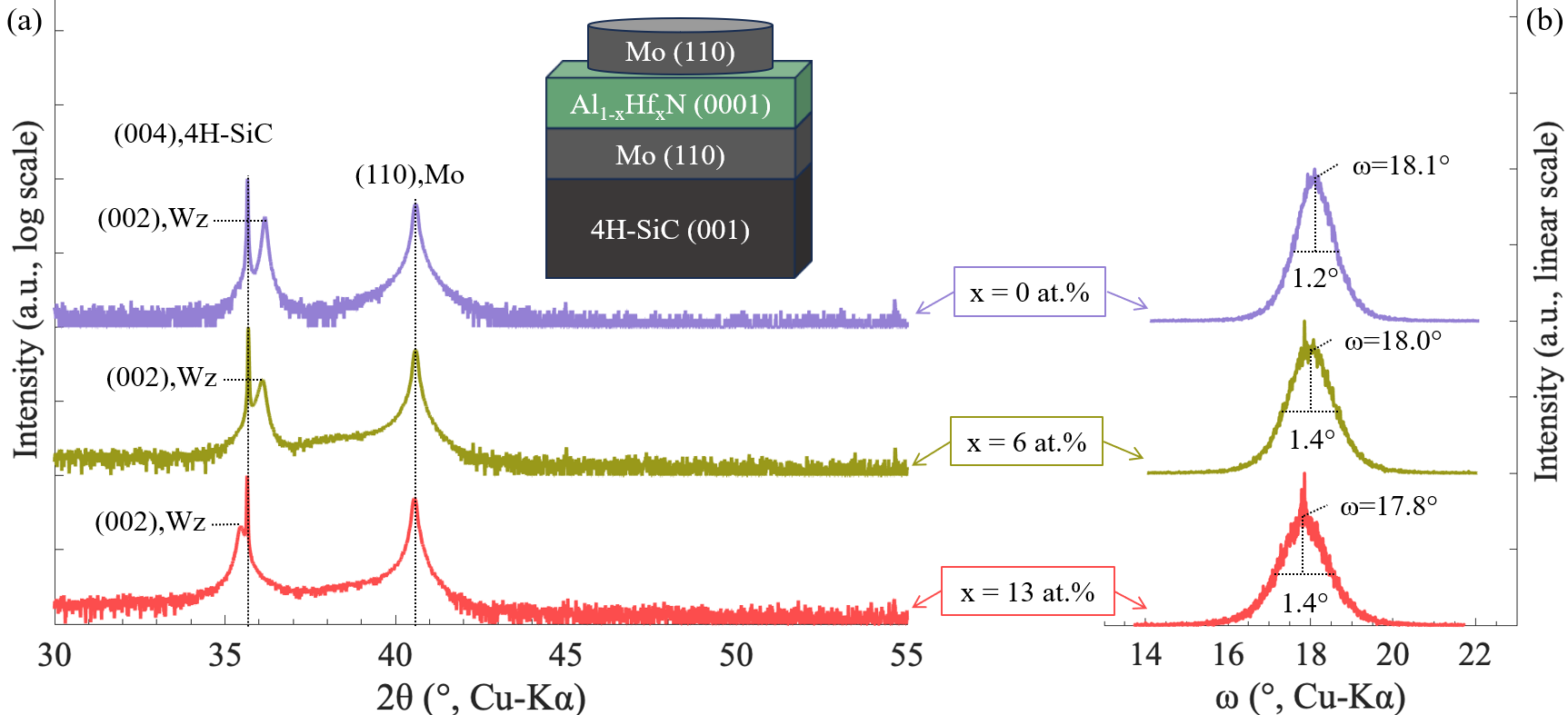}
\caption{\label{fig:XRD} X-ray diffraction (a) $\uptheta$-2$\uptheta$ scans show peaks from the substrate, electrode, and (002)-Wz exclusively, indicating the films were single phase wurtzite. (b) The $\omega$-rocking curves show the films were well textured. Hf cation atomic percentages were measured with RBS.}
\end{figure*}

Figure~\ref{fig:XRD}(a) shows $\omega$-rocking curves and $\uptheta$-2$\uptheta$ scans (using a Panalytical X'Pert3 MRD XL diffractometer) on c-axis textured single-phase wurtzite films. No diffraction peaks originating from anomalously oriented grains (AOGs) were observed. Fig.~\ref{fig:XRD}(b) shows a full width half max (FWHM) range of 1.2° to 1.4°, which is similar to --- or better than --- previously reported FWHM values for sputtered ferroelectric nitrides.\cite{drury_understanding_2021, nie_characterization_2022, suceava_enhancement_2023} Additional diffraction data, including lattice parameters, are shown in the Supporting Material. 

\begin{figure*}
\centering
\includegraphics[scale=0.8]{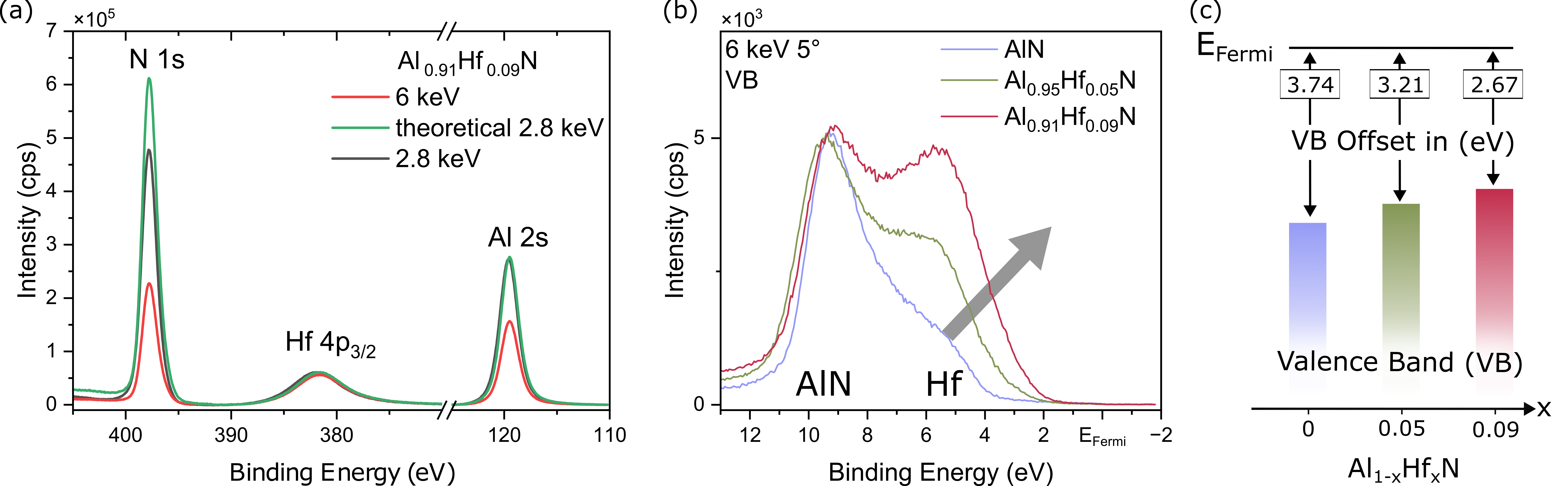}
\caption{\label{fig:HAXPES} HAXPES measurement of Al$_{1-x}$Hf$_x$N films. (a) Core level spectra of Al$_{0.91}$Hf$_{0.09}$N measured at 6 keV and 2.8 keV indicating a homogeneous Al:Hf distribution. (b),(c) Valence band spectra measured at 6 keV, comparing AlN, Al$_{0.95}$Hf$_{0.05}$N and Al$_{0.91}$Hf$_{0.09}$N. A reduced valence band offset (VBO) is observed with increasing Hf incorporation, however, a non-metallic character is maintained for all Hf concentrations.}
\end{figure*}

Figure~\ref{fig:HAXPES} shows valence band electronic structure determined by hard X-ray photoelectron spectroscopy (HAXPES)\cite{Mueller2022, Rehm2024} of Al$_{1-x}$Hf$_x$N films with varying stoichiometry (nominally $x$ = 0.05 and 0.09). Fig.~\ref{fig:HAXPES}(a) shows core-level spectra of N 1s, Hf 4p$_{3/2}$, and Al 2s acquired at 6 keV (bulk-sensitive to a depth of $\sim$18 nm) and 2.8 keV (surface-sensitive to a depth of $\sim$9 nm). The strong agreement between the theoretical and measured 2.8 keV spectra for Hf 4p$_{3/2}$ and Al 2s suggests a homogeneous distribution of Hf and Al throughout the Al$_{1-x}$Hf$_x$N layer. The measured N 1s intensity at 2.8 keV is lower than predicted, indicating a nitrogen deficiency at the surface, likely due to oxidation in uncapped samples. Fig.~\ref{fig:HAXPES}(b) displays valence band (VB) spectra of Al$_{1-x}$Hf$_x$ N films with different Hf contents, measured using 6 keV photons. Increasing Hf content results in a noticeable shift of the valence band maximum towards the Fermi level, reducing the valence band offset (VBO) relative to AlN from 3.7 eV to 2.7 eV for Al$_{0.91}$Hf$_{0.09}$N, as schematically depicted in Fig.~\ref{fig:HAXPES}(c). A non-metallic character is indicated for all three Al$_{1-x}$Hf$_x$N samples. Additional details on HAXPES experiments and data analysis are in the Supporting Material.

\begin{figure}
\includegraphics[scale=0.25]{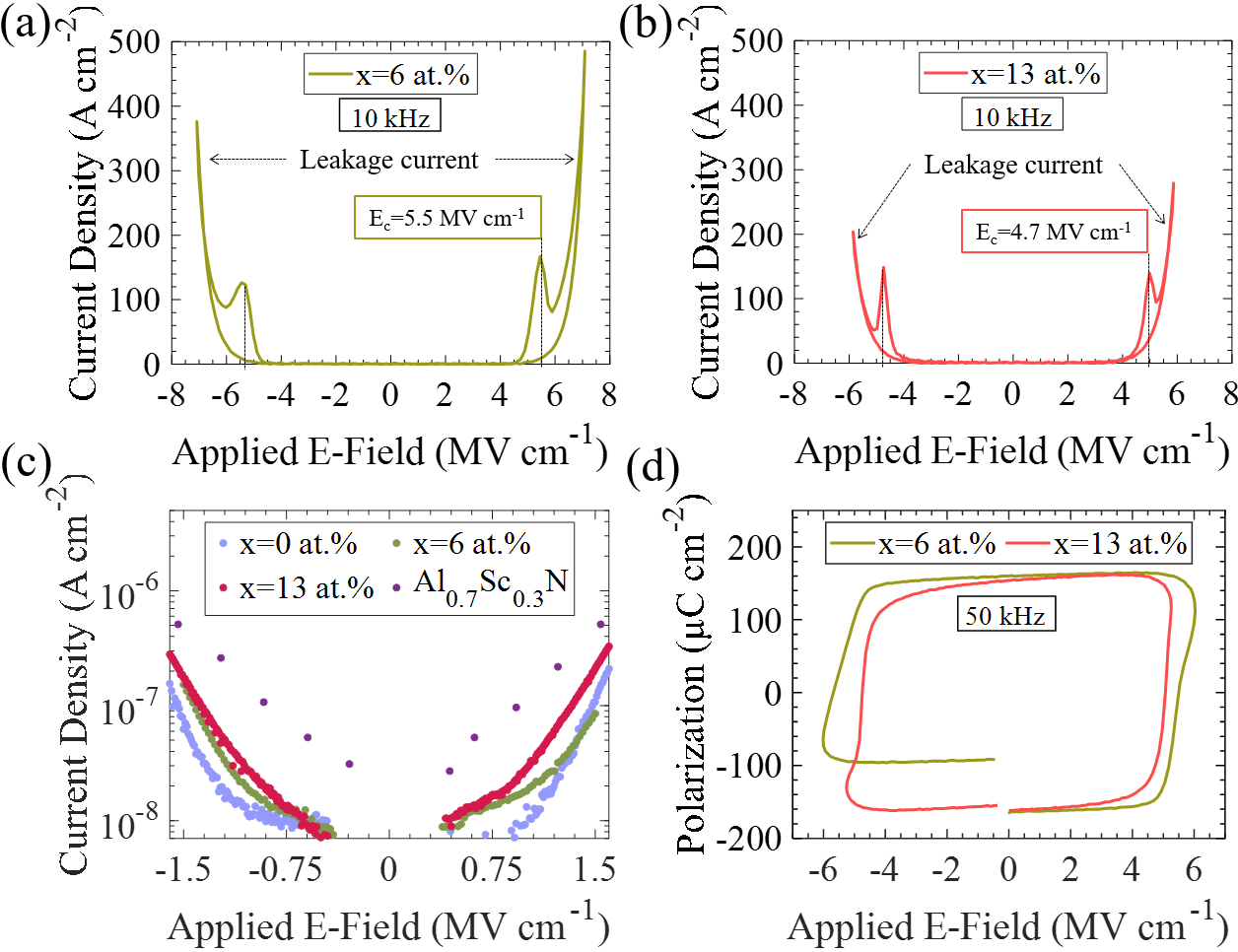}
\caption{\label{fig:elec_meas} Electrical measurements showed FE switching for films with $x$ = 0.06 and $x$ = 0.13. (a)(b) Hysteresis measurements of current density vs.\ applied E-field showed FE switching for films with $x$ = 0.06 \& 0.13 respectively. (c) These $x$ = 0.06 \& 0.13 films were insulators and had current densities at low E-field that were less than comparable Al\textsubscript{0.7}Sc\textsubscript{0.3}N samples.\cite{drury_high-temperature_2022} (d) Polarization vs.\ applied E-field loops at 50 kHz.}
\end{figure}

Figure~\ref{fig:elec_meas} shows the results of electrical measurements. For all measurements in this letter, a positive E-field direction is defined as pointing from the bottom electrode to the top electrode, normal to the substrate. Figures ~\ref{fig:elec_meas}(a) and ~\ref{fig:elec_meas}(b) show unambiguous ferroelectric switching current peaks. Data in Figures \ref{fig:elec_meas}(a) \& \ref{fig:elec_meas}(b) were collected by applying a triangular wave to the samples at 10 kHz, using a  tester from Radiant Technologies, Inc. From these data, a value of coercive electric field (E\textsubscript{c}) was extracted at $\pm$5.5 MV/cm \& $\pm$4.7 MV/cm for $x$ = 0.06 \& 0.13, respectively, based on the switching current peak. These data showed the same trend of E\textsubscript{c} vs.\ $x$ as other ferroelectric nitrides: increasing \textit{M}-element atomic fraction results in decreased E\textsubscript{c}.\cite{fichtner_alscn_2019,hayden_ferroelectricity_2021,lee_prediction_2024,lee_switching_2024,yazawa_local_2022} 
Fig.~\ref{fig:elec_meas}(c) shows that at low applied E-field, the $x$ = 0.06 \& 0.13 samples were better insulators than typical ferroelectric Al\textsubscript{0.7}Sc\textsubscript{0.3}N films\cite{drury_high-temperature_2022}. Data in Fig.~\ref{fig:elec_meas}(c) were taken using a Keithley 4200A-SCS. Polarization vs.\ applied E-field (P-E) loops are shown in \ref{fig:elec_meas}(d) at 50 kHz (see the Supplementary Material for 10 kHz loops). The top electrode diameter used was 50 $\mu$m for all measurements shown in Fig.~\ref{fig:elec_meas} and was connected to the signal ground for all tests.

\begin{figure*}
\centering
\includegraphics[scale=0.35]{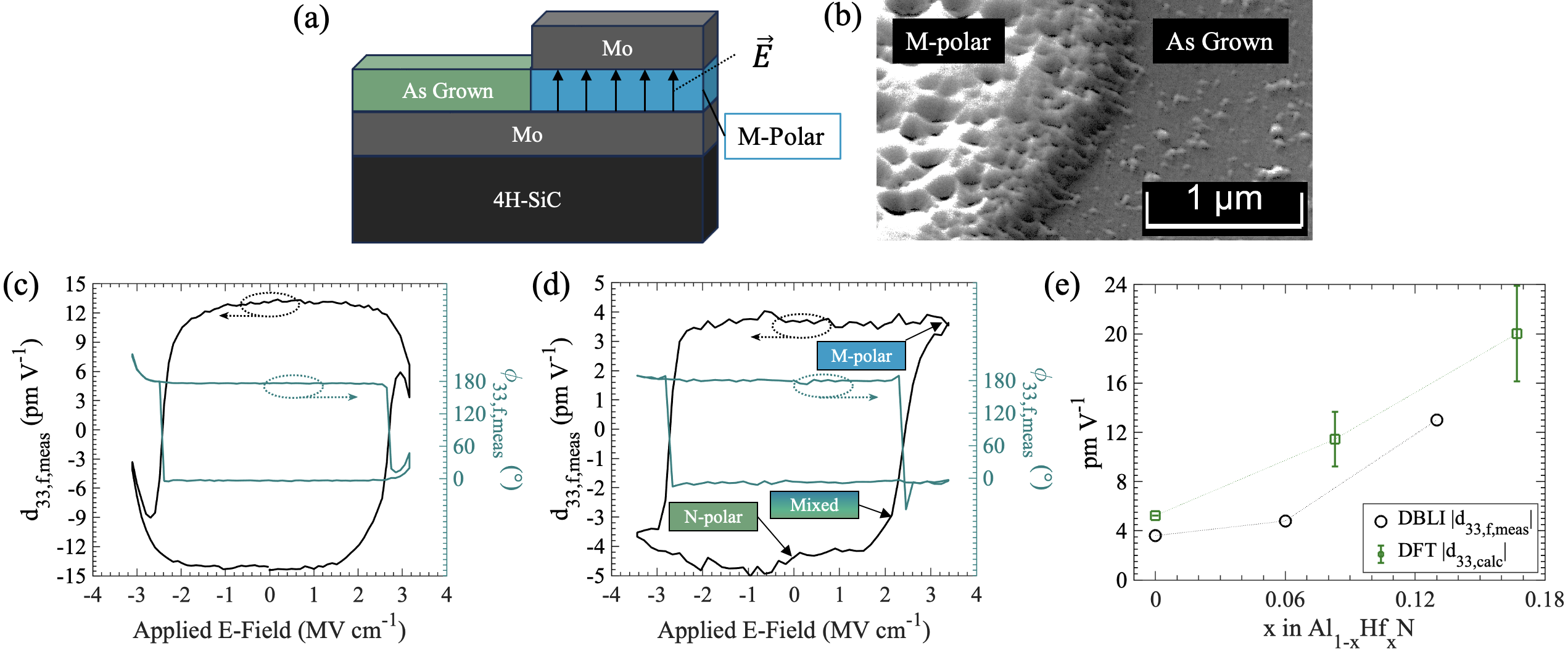}
\caption{\label{fig:pol_inver} Polarity inversion confirmed using acid etching and piezoelectric measurements. (a) Film pulsed M-polar. (b) SEM image of the M-polar step edge after etching in 80°C H\textsubscript{3}PO\textsubscript{4}. (c) DBLI piezoelectric measurement for $x$ = 0.13 film. $\upphi_{33,f,meas}$ deviating from 0° \& 180° is hypothesized to be due to leakage current. (d) DBLI measurement for x=0.06 with markers showing the change in polarity of the film around the ``staircase'' loop. See SM for more details. (e) Magnitude of the DBLI-measured (|d$_{33,f,meas}$|) and DFT-calculated (|d$_{33,calc}$|) piezoelectric coefficient for zero E-field. Connecting lines between points are a visual aid. (c)(d)(e) |d$_{33,f,meas}$| data were not compensated for substrate clamping effects of the 4H-SiC substrate. The loop period of (c)(d) was 800 seconds, leading to a much lower E\textsubscript{c} than shown in Figures~\ref{fig:elec_meas}(a) and \ref{fig:elec_meas}(b)\cite{scott_models_1996}}
\end{figure*}

To support the claim that the current density peaks in Figures \ref{fig:elec_meas}(a) \& \ref{fig:elec_meas}(b) were due to ferroelectricity, further measurements were carried out using small signal d$_{33}$ testing. The first part of this measurement employed multiple voltage pulses applied to the Mo top electrodes on the $x$ = 0.06 \& 0.13 films to ensure that the film region beneath the electrodes was fully switched into the N-polar state. Then, a small signal sinusoidal E-field was applied to the films with a frequency of 200 Hz and an amplitude of 0.5 MV/cm (E\textsubscript{ss}). The DBLI was then used to obtain the as-measured value of the piezoelectric coefficient: d$_{33,f,meas}$. The magnitude of d$_{33,f,meas}$ is shown in Fig.~\ref{fig:pol_inver}(e). Fig.~\ref{fig:pol_inver}(e) also shows d$_{33,calc}$ values calculated using density functional theory (DFT). The d$_{33,f,meas}$ values increase up to 12 pm/V at $x$ = 0.13, and the qualitative trend for calculation and measurement matched. Note that the absolute value is not quantitatively comparable between DFT and experimental results since DFT calculations are performed assuming single crystals, and d$_{33,f,meas}$ is affected by clamping.\cite{sivaramakrishnan_concurrent_2018}

After the switching pulse and subsequent measurement of d$_{33,f,meas}$, a bias electric field was applied to the films as a staircase function superimposed with E\textsubscript{ss} (see the Supporting Materials for an illustration of the function). Figures~\ref{fig:pol_inver}(c) and ~\ref{fig:pol_inver}(d) show the result of this measurement for films with $x$ = 0.13 \& 0.06, respectively. The respective films switched from N-polar to M-polar upon exceeding the bias field needed to switch, as illustrated by labels indicating film polarity in Fig.~\ref{fig:pol_inver}(d). The magnitude of each measurement step was 0.1 MV/cm. The dwell time after each step in voltage was 5 seconds allowing for 1000 averages of E\textsubscript{ss} to occur. This averaging was necessary for extracting the angstrom-level field induced displacement. A total of 160 points were collected and therefore the loop period of the measurements in Figures~\ref{fig:pol_inver}(c) and ~\ref{fig:pol_inver}(d) was 800s. 

The reduced E\textsubscript{c} observed in Fig.~\ref{fig:pol_inver} (c) and \ref{fig:pol_inver}(d) relative to Fig.~\ref{fig:elec_meas}(b) is attributed to the increased applied E-field period of the measurement:\cite{scott_models_1996} 100µs for Fig.~\ref{fig:elec_meas}(b) versus 800s for Figures ~\ref{fig:pol_inver}(c) and ~\ref{fig:pol_inver}(d). As the applied bias E-field was cycled, an unambiguous change occurred in the sign of d$_{33,f,meas}$ and $\upphi_{33,f,meas}$ shifted from 0° to 180°, indicating polarization inversion occurred. The applied E-field loops shown in Figures \ref{fig:pol_inver}(c) and \ref{fig:pol_inver}(d) corroborate the ferroelectric behavior of these Al$_{1-x}$Hf$_{x}$N films. The decrease in d$_{33,f,meas}$ when |E| > 3MV/cm and $\upphi_{33,f,meas}$ deviating from 0° \& 180° seen in Fig.~\ref{fig:pol_inver}(c) is hypothesized to be due to an increased real component magnitude of the impedance of the sample.

Anisotropic acid etching further confirmed that field-induced polarization reversal occurred. H$_{3}$PO$_{4}$ was applied to a film where a region had been pulsed metal polar (see Fig.~\ref{fig:pol_inver}(a)). The step edge of the remnant M-polar region can be seen in Fig.~\ref{fig:pol_inver}(b) while the surrounding as-grown N-polar region was dissolved, since H$_{3}$PO$_{4}$ dissolves N-polar surfaces faster than M-polar surfaces.\cite{zhuang_wet_2005,fichtner_alscn_2019}

Using DFT, we calculated d$_{33}$ as a function of Hf context for Al$_{1-x}$Hf$_{x}$N alloys (see SM for details). Fig.~\ref{fig:pol_inver}(e) shows that the predicted value for pure AlN (5.22 pm/V) is consistent with previous DFT-based prediction ($\sim$4-5 pm/V).\cite{iwazaki_highly_2015,startt_unlocking_2023} Overall, Hf substitution with charge compensation by Al vacancies increases d$_{33}$, and the trend of d$_{33}$ vs.\ $x$ qualitatively agrees with the experimental trend, noting again the differences in single crystal calculations versus clamped thin film measurements.

\begin{figure}
\centering
\includegraphics[width=\linewidth]{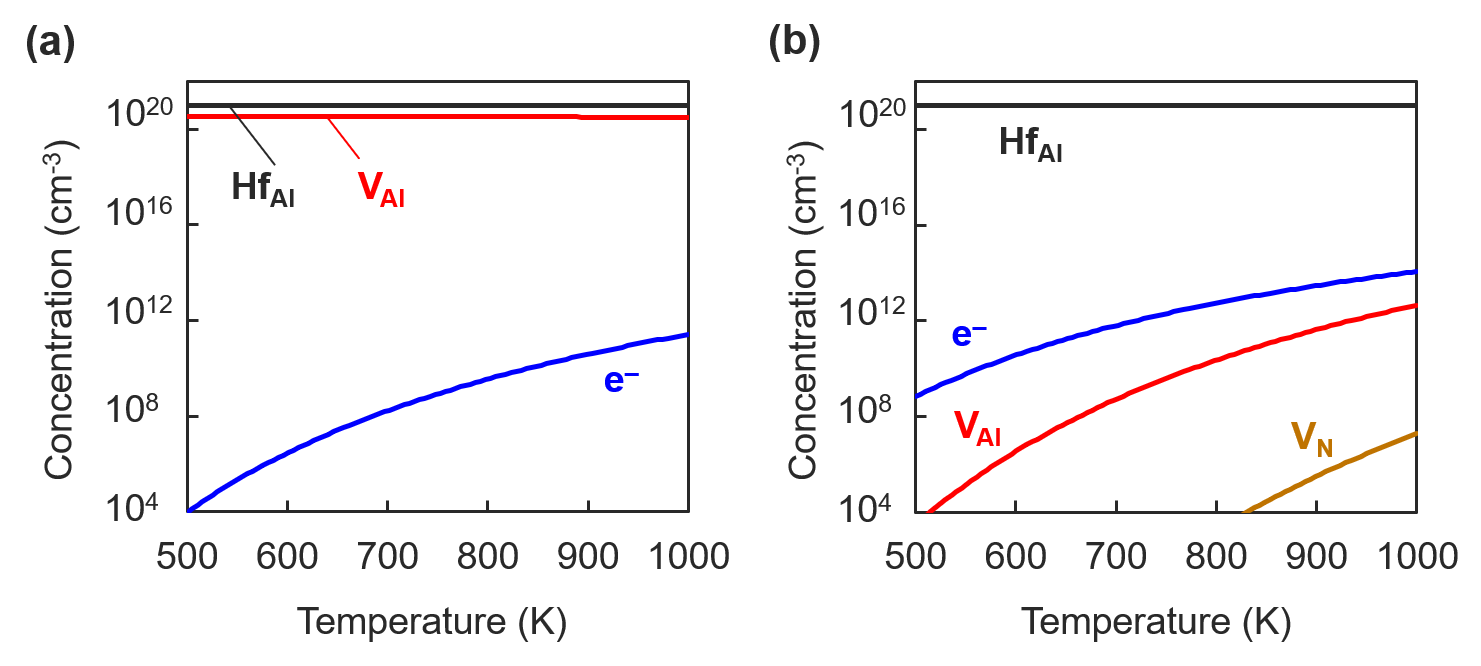}
\caption{\label{fig:dft} Carrier and defect concentrations as a function of temperature under (a) N-rich and (b) Al-rich growth conditions.}
\end{figure}

To clarify the charge compensation mechanism for the typically tetravalent dopant Hf, we calculated defect and carrier concentrations at constant Hf concentration of $10^{21}/\mathrm{cm^3}$ (= 2.1 at.~\%) as shown in Fig.\ref{fig:dft} (see SM for details). Under N-rich growth conditions, $V_{\mathrm{Al}}^{'''}$ exhibit a low formation energy as described in Fig.~S7(a). Consequently, the resultant high concentration of $V_{\mathrm{Al}}^{'''}$ almost compensate the positive charge of $\mathrm{Hf}_{\mathrm{Al}}^{\bullet}$ [see Fig.~\ref{fig:dft}(a)]. On the other hand, the high formation energy of $V_{\mathrm{Al}}^{'''}$ under Al-rich growth conditions [Fig.~S7(b)] leads to a lower concentration of $V_{\mathrm{Al}}^{'''}$, thus resulting in a higher electronic carrier concentration when compared with the N-rich conditions [Fig.~\ref{fig:dft}(b)]. The actual experimental conditions are considered to lie between these N-rich and Al-rich extremes. Therefore, our calculations suggest that the charges carried by Hf$_{Al}$ were partially (mostly) compensated by $V_{\mathrm{Al}}^{'''}$, leading to a relatively low electronic carrier concentration and overall insulating character of the Al$_{1-x}$Hf$_{x}$N films.

In summary, this work showed that Al$_{1-x}$Hf$_{x}$N sputtered thin films remain insulating and can be ferroelectric. This was shown via switching current peaks, small signal d$_{33,f,meas}$ DBLI measurements, and anisotropic etching to confirm polarization inversion. DFT calculations and HAXPES experiments both reinforced the electrical measurement findings that the material system is electrically insulating. While this work did not focus on HOT-NVM or FeHEMTs specifically, this work does enable researchers to think beyond trivalent substitutions when designing devices, opening up more possibilities for better device performance. Future work is planned to identify and quantify specific charge compensation mechanisms and electronic structure for Al$_{1-x}$Hf$_x$N with higher fidelity. 

\begin{acknowledgments}
This work was co-authored by Colorado School of Mines and the National Renewable Energy Laboratory, operated by the Alliance for Sustainable Energy, LLC, for the U.S. Department of Energy (DOE) under Contract No. DE-AC36-08GO28308. A portion of this work was supported by the NSF DMREF program award DMR-2119281. Some of the work was performed in following core facility, which is a part of Colorado School of Mines’ Shared Instrumentation Facility (SCR\_022047, SCR\_022047). A portion of the research was sponsored by the Army Research Laboratory and was accomplished under Cooperative Agreements W911NF-21-2-0210 and W911NF-19-2-0119. Thanks to Ande Bryan, Eli Cooper, Alex Dixon, Nastazia Moshirfatemi, and Margaret Brown for technical assistance with film fabrication and/or testing. The research was performed using computational resources sponsored by the Department of Energy's Office of Energy Efficiency and Renewable Energy and located at NREL. The views and conclusions contained in this document are those of the authors and should not be interpreted as representing the official policies, either expressed or implied, of the DEVCOM Army Research Laboratory, DOE, or the U.S.\ Government. The U.S.\ Government is authorized to reproduce and distribute reprints for Government purposes notwithstanding any copyright notation herein. MM and OR acknowledge support by the Deutsche Forschungsgemeinschaft through Sonderforschungsbereich SFB 1432 (Project No. 425217212, Subproject No. A07), by University of Konstanz BlueSky initiative, and by the VECTOR Foundation (project iOSMEMO). We acknowledge DESY (Hamburg, Germany), a member of the Helmholtz Association HGF, for the provision of experimental facilities. Beamtime was allocated for proposal R-20240665. Funding for the HAXPES instrument at beamline P22 by the Federal Ministry of Education and Research (BMBF) under contracts 05KS7UM1 and 05K10UMA with Universität Mainz; 05KS7WW3, 05K10WW1, and 05K13WW1 with Universität Würzburg is gratefully acknowledged.

\end{acknowledgments}

\bibliographystyle{apsrev4-1}
\bibliography{AlHfN_APL}

\end{document}



\title{Supporting Material for Ferroelectricity of Wurtzite Al\textsubscript{1-x}Hf\textsubscript{x}N Heterovalent Alloys} 



\author{Nate S.P. Bernstein}
 \affiliation{Department of Metallurgical and Materials Engineering, Colorado School of Mines, Golden, Colorado 80401, United States}
 \email{nbernstein@mines.edu}
\author{Daniel Drury}
 \affiliation{U.S. Army Combat Capabilities Development Command - Army Research Laboratory, Adelphi, Maryland 20783, United States}
\author{Cheng-Wei Lee}
 \affiliation{Department of Metallurgical and Materials Engineering, Colorado School of Mines, Golden, Colorado 80401, United States}
 \affiliation{Materials Science Center, National Renewable Energy Laboratory, Golden, Colorado 80401, United States}
\author{Tatau Shimada}
 \affiliation{TAIYO YUDEN CO., LTD., 2-7-19, Kyobashi, Chuo-ku, Tokyo, 104-0031, Japan}
\author{Yuki Sakai}
 \affiliation{TAIYO YUDEN CO., LTD., 2-7-19, Kyobashi, Chuo-ku, Tokyo, 104-0031, Japan}
 \author{Oliver Rehm}
 \affiliation{Department of Physics, Universit\"{a}t Konstanz, D-78457 Konstanz, Germany}
\author{Lutz Baumgarten}
 \affiliation{Forschungszentrum J\"{u}lich GmbH, Peter Grünberg Institut (PGI-6), D-52425 J\"{u}lich, Germany}
\author{Martina M\"{u}ller}
 \affiliation{Department of Physics, Universit\"{a}t Konstanz, D-78457 Konstanz, Germany}
\author{Prashun Gorai}
 \affiliation{Department of Chemical and Biological Engineering, Rensselaer Polytechnic Institute, Troy, New York 12180, United States}
 \affiliation{Department of Metallurgical and Materials Engineering, Colorado School of Mines, Golden, Colorado 80401, United States}
\author{Yoshiki Iwazaki}
 \affiliation{TAIYO YUDEN CO., LTD., 2-7-19, Kyobashi, Chuo-ku, Tokyo, 104-0031, Japan}
\author{Glen R. Fox}
 \affiliation{Fox Materials Consulting, LLC, Colorado Springs, Colorado 80908, United States}
\author{Keisuke Yazawa}
 \affiliation{Department of Metallurgical and Materials Engineering, Colorado School of Mines, Golden, Colorado 80401, United States}
 \affiliation{Materials Science Center, National Renewable Energy Laboratory, Golden, Colorado 80401, United States}
\author{Brendan Hanrahan}
 \affiliation{U.S. Army Combat Capabilities Development Command, Adelphi, Maryland 20783, United States}
\author{Geoff L. Brennecka}
 \affiliation{Department of Metallurgical and Materials Engineering, Colorado School of Mines, Golden, Colorado 80401, United States}
 \email{gbrennec@mines.edu}
 
\date{\today}

\pacs{}

\maketitle 

\renewcommand{\thefigure}{S\arabic{figure}} 
\begin{figure*}[hbt!]
\includegraphics[scale=0.5]{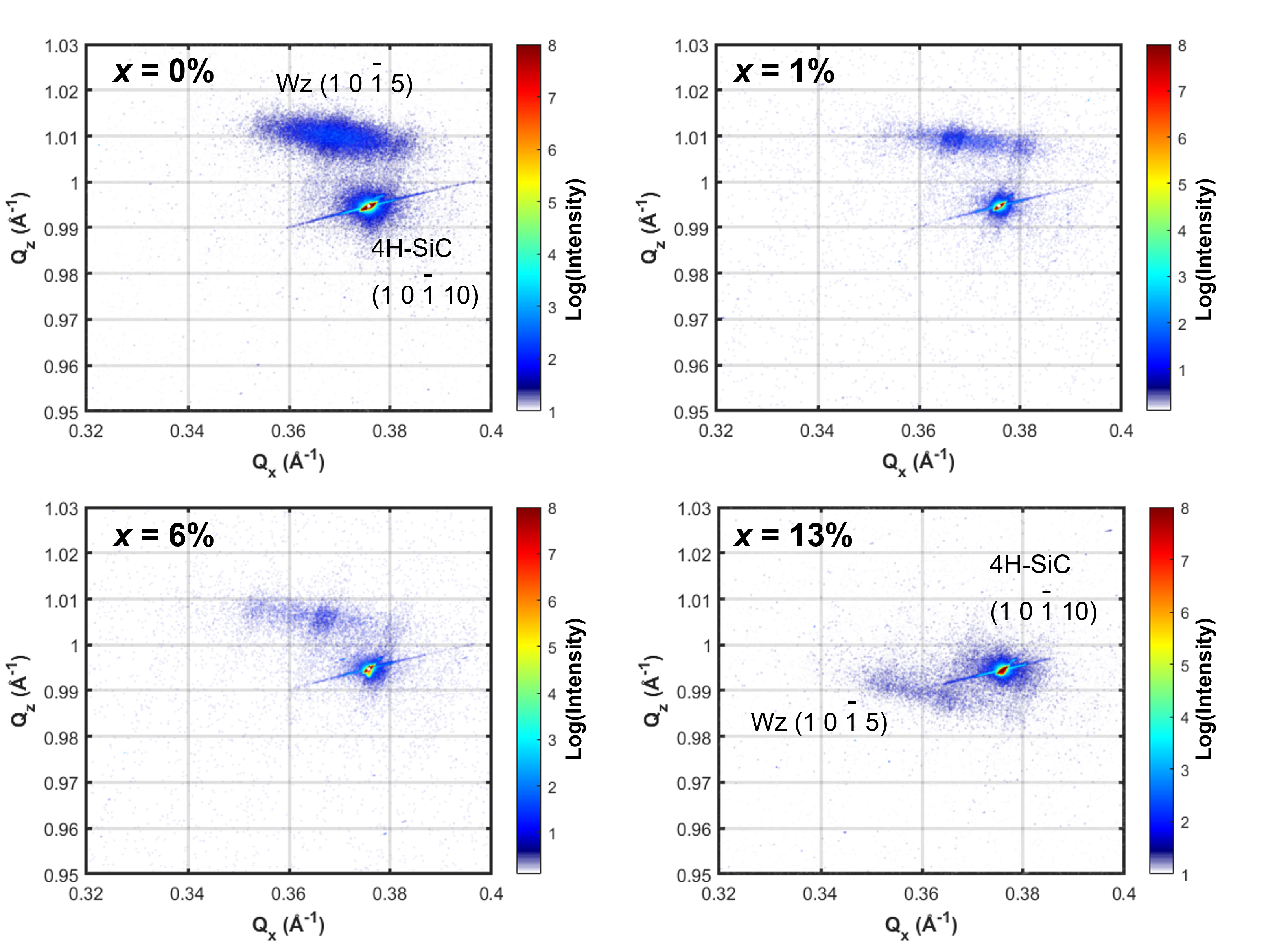}
\caption{\label{fig:SM_XRD} Reciprocal Space Maps of Al$_{1-x}$Hf$_x$N/(110)Mo/(001)SiC.}
\end{figure*}

\begin{figure}[hbt!]
    \centering
    \includegraphics[width=1\linewidth]{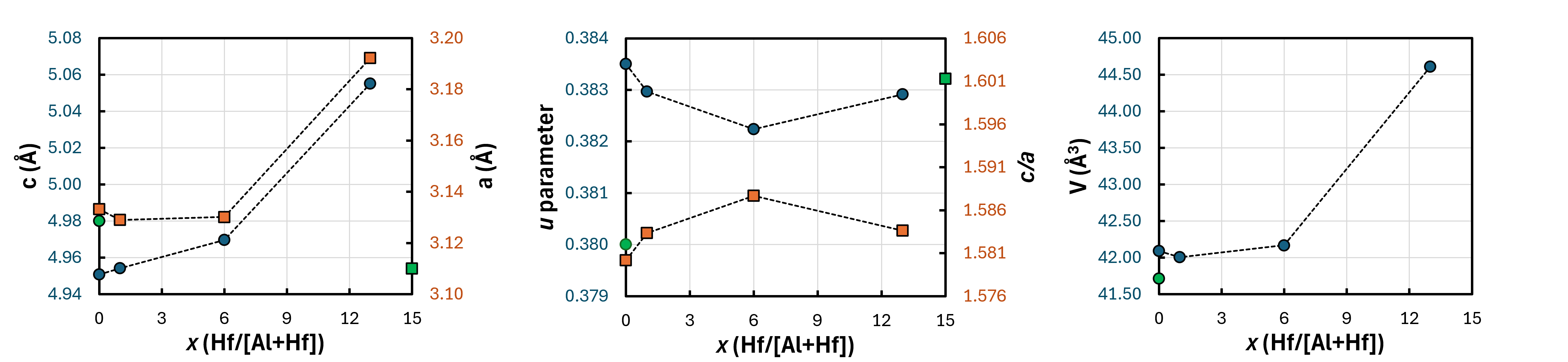}
    \caption{Lattice parameters, $u$ parameters\cite{schulz_crystal_1977}, and unit cell volume for Al$_{1-x}$Hf$_x$N/(110)Mo/(001)SiC.}
    \label{fig:SM_lattice_parameters}
\end{figure}

\begin{figure*}[hbt!]
\includegraphics[scale=0.5]{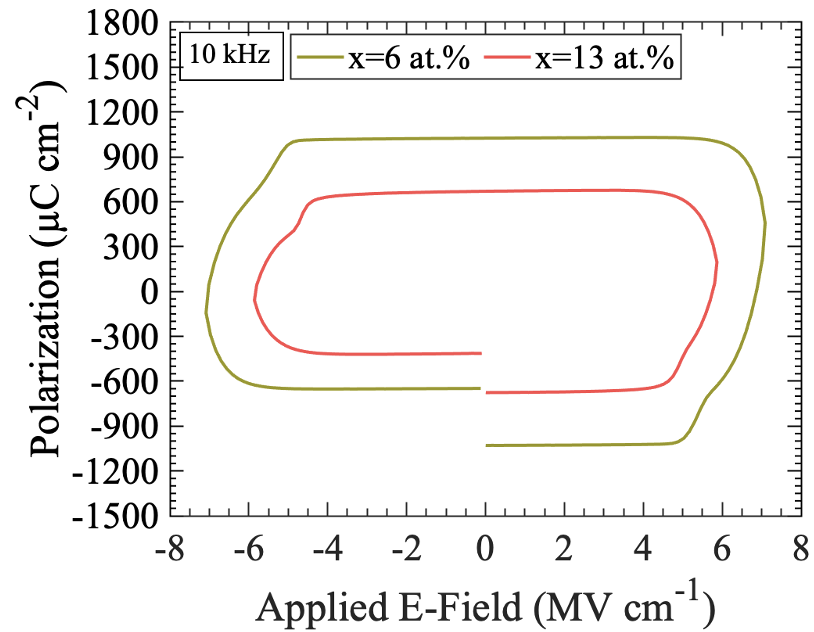}
\caption{\label{fig:SM_PE_loops} P-E loop corresponding to Figures 2(a) \& 2(b)}
\end{figure*}

\begin{figure*}[hbt!]
\includegraphics[scale=0.5]{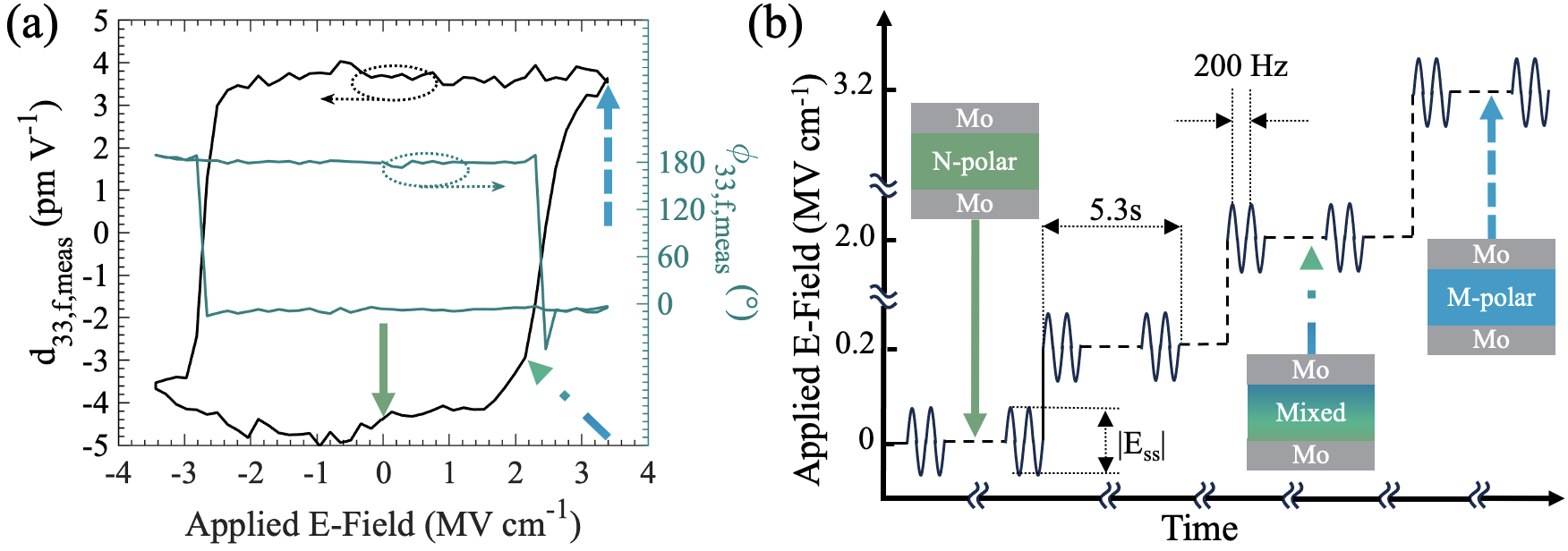}
\caption{\label{fig:SM_pol_inver} Polarity inversion shown via DBLI diplacement measurements and an illustration of the staircase function used for these measurements. (a) DBLI-measured effective piezoelectric charge coefficients for x=0.06 with markers mapping to (b). (b) Illustration of applied electric field for small signal d\textsubscript{33,f,meas} measurements.}
\end{figure*}

\begin{figure*}[hbt!]
\includegraphics[scale=0.45]{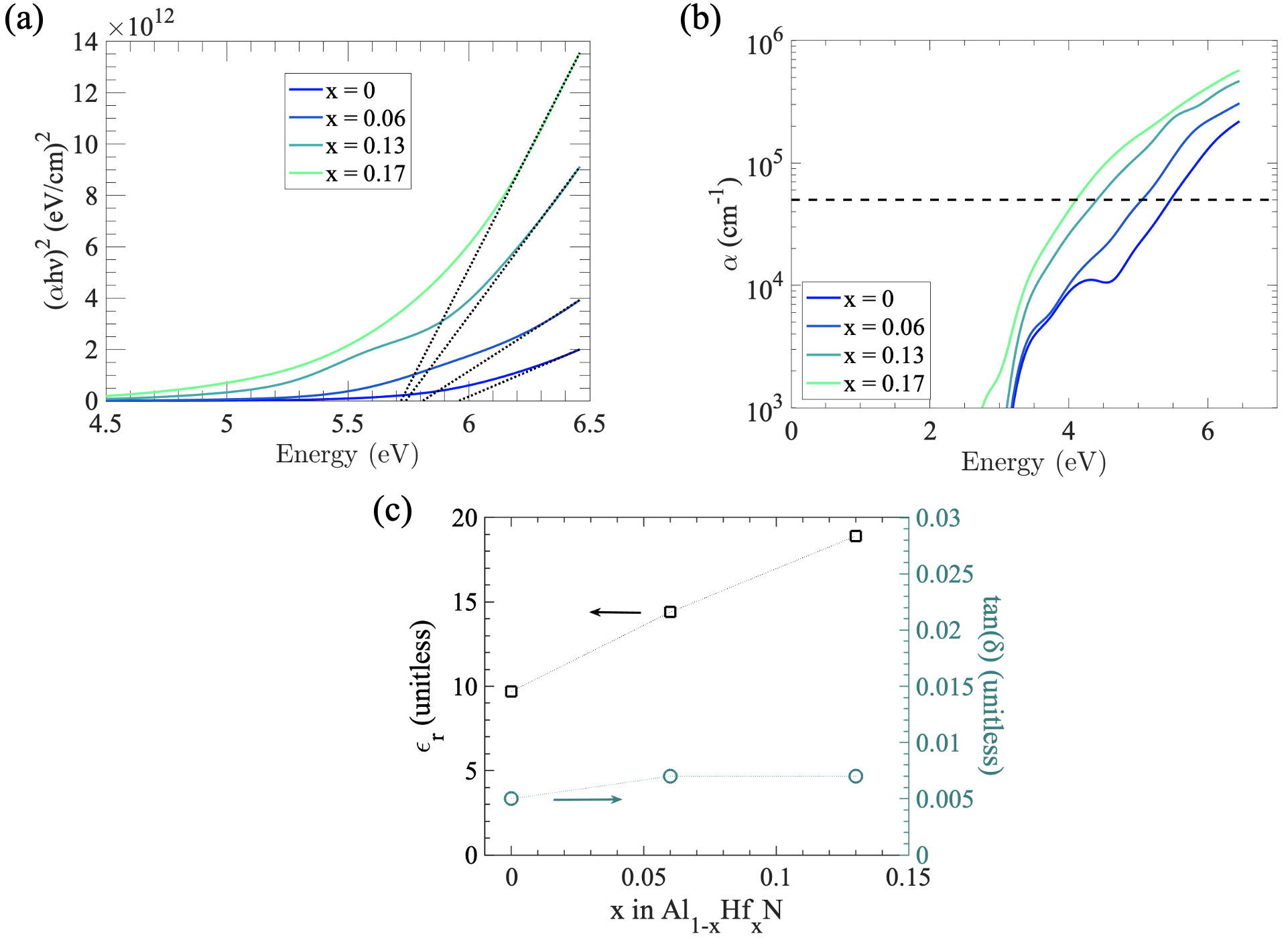}
\caption{\label{fig:SM_dft_ellip} The material system has optical properties indicative of an electrical insulator. (a) Tauc plot with maximum energy insufficiently larger than the band gap to extract the exact bandgap for x<0.17. Regardless, the plot clearly shows the E\textsubscript{g}>5eV, supporting electrical measurement results that showed the material system is an insulator. (b) Absorption onset for E$\upalpha$=5$\times 10^4$ cm\textsuperscript{-1} showing optical absorption does not  occur until >4eV for all samples. Similar ripple has been seen in other AlN measurements \cite{aita_optical_1989,demiryont_optical_1986} (c) Electrodes measured on the Mo//SiC samples have a loss tangent and electrical permittivity typical of a dielectric, also further reinforcing the insulator nature of the material system.}
\end{figure*}

\newpage
\section{Experimental Methods}
\subsection{Mo/Al\textsubscript{1-x}Hf\textsubscript{x}N/Mo/SiC Synthesis}
Substrate preparation was done in an ISO 5 (Class 100) clean room and synthesis was  done in an  ISO 4 (Class 10) clean room. Dicing was done with a diamond saw. The deposition side of the wafer was protected using Shipley microposit FSC photoresist. The not-deposited-on side was held in place with Semicorp standard dicing tape. After dicing, wafers were cleaned using acetone sonication for 5 minutes, and removed without allowing the acetone to evaporate on the surface of the film. After removal from acetone, the sample was rinsed with methanol, then isopropanol, then blown with dry N$_2$ (AMI rinse). Post AMI rinse, the samples were submerged in buffered oxide etchant (BOE 6:1) for 60 seconds to remove the oxide layer on the surface of the SiC, which was followed by a DI rinse and dried with N$_2$. Care was taken not to touch the surface of the substrate and substrates were held with tweezers on the side of the die.

Prior to any depositions, targets used were presputtered for 3 hours at 2 mTorr and 40 sccm of ultra-high purity argon (UHP Ar). Substrates were loaded into the chamber via a load lock. A base pressure of <8$\times$10\textsuperscript{-8} torr was reached after loading the sample into the process chamber. The substrate heater was then programmed to a 550 °C set point. The Mo bottom electrode was DC magnetron sputtered with a power density of 59 W/cm$^2$ and a working pressure of 2 mTorr with 40 sccm of UHP Ar flowing. Prior to deposition of Mo on the substrate, 4 minutes of presputtering was done on the Mo target under the same deposition conditions of 2 mTorr and 40 sccm UHP Ar, with a gun shutter closed.

The nitride layer was sputtered with the following conditions: 2 mTorr of Ar/N$_2$ (40/40 sccm flow), a substrate heater setpoint of 400 °C, and a sputtering time of 100 min. This ratio of Ar to N$_2$ set the sputtering mode to  metallic target mode, as opposed to a poisoned (or nitridized) mode. Prior to deposition of Al and Hf on the substrate, 4 minutes of presputtering was done on the targets under the same 2 mTorr and Ar/N$_2$ (40/40) conditions, with gun shutters closed. A pulsed DC power source and an RF source were used to sputter the target Al and Hf targets, respectively. A target to substrate distance of 10 cm was used. Sputter guns were at a 45° angle relative to the substrate. The Al target power density was a constant 59 W/cm$^2$ for all three samples. The pulsed DC supply used a 100 kHz frequency and 1.5 $\mu$s pulse duration.

After the nitride layers were deposited, vacuum was broken, and a photolithographic process was used to create top electrodes. Photoresist was spun onto the films and then developed using a mask \& UV lamp. TMAH was used as a developer for the electrode pattern. The top Mo layer was DC magnetron sputtered with the same 59 W/cm$^2$ \& 2 mTorr Ar as the bottom Mo layer at ambient temperature. Liftoff in acetone sonication was done after Mo deposition on top of the photoresist.

\subsection{Ellipsometry Measurement}
Ellipsometry measurements shown in Fig.~\ref{fig:SM_dft_ellip} were done to quantify the absorption coefficient of the films. The samples used for this measurement were grown on native oxide silicon instead of Silicon Carbide, to make the modeling of the optical properties of the films simpler. The ellipsometer used was a Wollam M-2000 with a lamp capable of outputting a maximum energy of 6.45 eV. The lamp was turned on for 30 minutes prior to making a measurement to allow the lamp to reach thermal equilibrium. The ellipsometry measurement was taken at 70° and 80° to capture data at the Brewster angle of a thin film on silicon, maximizing the precision of the ellipsometry measurement. After the measurement was taken, two different analyses were done on the data: (1) A Cauchy relationship (2) B-spline fitting. The Cauchy relationship was fit to data only up to 2.5 eV. An energy of 2.5 eV was sufficiently low as to capture the transparent (non-absorbing) region of the data. The Cauchy model was used to obtain the film thickness and surface roughness. Next, B-spline fitting was used, with the thickness \& roughness obtained from the Cauchy model, to obtain the absorption coefficient of the samples measured. Both the Cauchy model and the B-spline fitting routine used the JA Wollam model for native oxide silicon.

\subsection{Al\textsubscript{1-x}Hf\textsubscript{x}N/Si Synthesis}
Thin films of Al$_{1-x}$Hf$_{x}$N used during ellipsometry measurements were deposited on native oxide silicon and wafers of undoped silicon. Substrate preparation was  done in an  ISO 4 (Class 10) clean room. The Si wafer was removed from the wafer carrier (a.k.a.\ front opening universal pod). Dicing was done by cleaving with a diamond scribe pen. Die were blown with dry N$_2$ to remove any silicon dust. Care was taken to avoid touching the native oxide surface of the silicon to avoid contamination.

The same chamber and sputtering targets were used to prepare the Al\textsubscript{1-x}Hf\textsubscript{x}N/Si samples as the Mo/Al\textsubscript{1-x}Hf\textsubscript{x}N/Mo/SiC samples. A base pressure of <1$\times$10\textsuperscript{-7} Torr was reached after loading the sample. The substrate heater was then programmed to a 400°C set point. The film was sputtered with the following conditions: 2 mTorr of Ar/N$_2$ (40/40 sccm flow), a substrate heater setpoint of 400°C, and a sputtering time varying such that the film thickness was 95±5 nm. A pulsed DC power source was used to sputter the target Al target. A target to substrate distance of 10 cm was used. Sputter guns were at a 45° angle relative to the substrate. The Al target power density was a constant 59 W/cm$^2$ for all samples.

\subsection{Acid Etching}
Two electrodes were selected to be pulsed with an E-field pointing antiparallel \& parallel to the film normal, resulting in the electrodes being Nitrogen-polar (N-polar) and Metal-polar (M-polar) respectively. The direction of the E-field for the M-polar electrode was pointing from the bottom electrode to the top electrode. Both electrodes were pulsed with a 6 MV/cm E-field strength. After being pulsed the top electrodes were dissolved in aqua regia. Next, the sample region including the pulsed electrodes was submerged in 80°C H\textsubscript{3}PO\textsubscript{4} until the as-grown majority of the film was visibly dissolved. The M-polar region remained largely intact while the N-polar region was completely dissolved, just as the as-grown film around it. The anisotropic etch rate of the film occurred because H\textsubscript{3}PO\textsubscript{4} dissolves metal polar surfaces faster than nitrogen polar surfaces.\cite{zhuang_wet_2005,fichtner_alscn_2019} This acid etching experiment confirmed polarization inversion occurred for both the x = 0.06 \& x = 0.13 samples, further verifying ferroelectricity in Al$_{1-x}$Hf$_{x}$N.

\subsection{DBLI}
Small signal piezoelectric coefficient measurements were obtained using an aixacct aixDBLI research line instrument. First a quartz reference was used to calibrate the instrument such that the measured d$_{33}$ matched the reference value. Next, the incident top and bottom beams were aligned to be concentric. The reference beam and measurement beam were then adjusted such that the center of an interference pattern between the two beams was centered on the detector of the instrument. Minor adjustments were made in the intensity between the reference beam and the measurement beam to account for the difference in reflectivity of the samples relative to the quartz reference.

The as-measured value of the piezoelectric coefficient (d$_{33,f,meas}$) is distinct from the unclamped intrinsic coefficient of the film (d$_{33}$). The difference being that d$_{33,f,meas}$ does not account for how the clamping effect of the substrate --- combined with the electrode diameter to substrate thickness ratio --- affects the measurement (due to localized substrate strain). The effects of the Poisson ratio on DBLI measurements when using an Si substrate are defined \cite{sivaramakrishnan_electrode_2013,sivaramakrishnan_concurrent_2018} however, future work is needed to quantify converting  d$_{33,f,meas}$ to d$_{33,f}$ and the intrinsic d$_{33}$ when using SiC substrates. DBLI measurements reported for the three samples were done on 250 µm diameter electrode and 500 µm thick 4H-SiC substrates. It's unknown if this 1:2 ratio causes d$_{33,f,meas}$ > d$_{33,f}$ or if this 1:2 ratio causes d$_{33,f,meas}$ < d$_{33,f}$. For the case of d$_{33,f,meas}$ < d$_{33,f}$, substrate clamping may explain the discrepancy between d$_{33,calc}$ and d$_{33,f,meas}$. See the publication by Sivaramakrishnan et al. for further details, especially equation (1) in their work.\cite{sivaramakrishnan_concurrent_2018}

\section{HAXPES}
Hard X-ray Photoelectron Spectroscopy (HAXPES) was performed at the P22 beamline of PETRA III (DESY, Hamburg)\cite{Schlueter2019} to investigate element-selective chemical properties. Core level spectra of Al 2s, Hf 4p$_{3/2}$, N 1s and the valence band (VB) were recorded at a photon energy of 2.8 keV and 6 keV, providing an information depth of 9 nm and 18 nm, respectively. The information depths were estimated using the Simulation of Electron Spectra for Surface Analysis (SESSA) software from the National Institute of Standards and Technology (NIST)\cite{Werner2017}. HAXPES is therefore more depth sensitive than conventional laboratory XPS (Al K alpha or Mg K alpha), which is typically in the range of a few nanometers.\cite{Mueller2022}  A SPECS PHOIBOS 225HV electron analyzer was used at an emission angle of 5$^{\circ}$ and a pass energy of 50 eV, resulting in an overall energy resolution of approximately 300 meV. For quantitative analysis, a theoretical 2.8 keV spectrum is compared to the measured 2.8 keV spectrum of an Al$_{0.88}$Hf$_{0.12}$N layer (see the manuscript). This theoretical spectrum was derived from the 6 keV measurement by accounting for differences in photon flux and photoionization cross-sections with changing from 6 keV to 2.8 keV in excitation energy\cite{Trzhaskovskaya_2018}.

\subsection{Sample Information}
\begin{figure}
    \centering
    \includegraphics[width=0.5\linewidth]{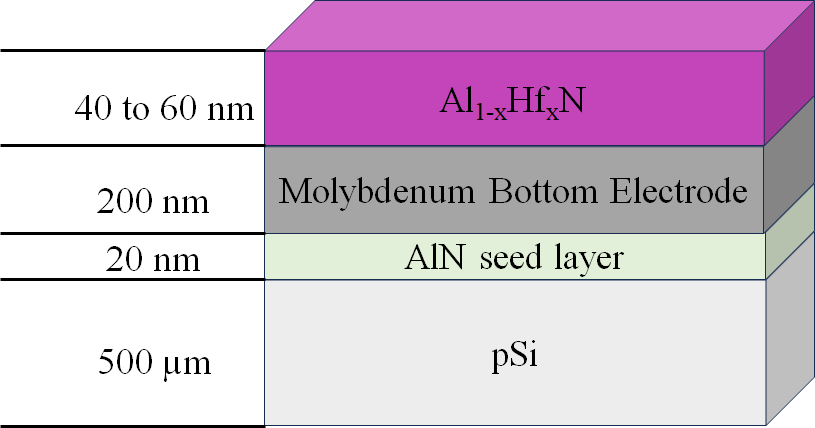}
    \caption{Stack for the three HAXPES samples}
    \label{fig:HAXPES_Stack}
\end{figure}

The stack for the three HAXPES samples is shown in figure \ref{fig:HAXPES_Stack}. 
Due to resource limitation the three samples synthesized for HAXPES testing were made in a different sputtering chamber than the ones synthesized for electrical testing. The Al target used was 99.9 \% purity (Kurt J. Lesker Co.) and the Hf target used was 99.9 \% purity (Stanford Advanced Materials). RF power sources were used for both targets. The target diameters were 5.08 cm. Target power density for the Al target was 64 W/cm$^2$. Target power density for the Hf target was 0, 29, and 48 W/cm$^2$. The sputter gun containing the Al target was at a 0° angle relative to the substrate (direct incidence) and had a target to substrate distance of \~10 cm. The sputter gun containing the Hf target was at an \~30° angle relative to the substrate and had a target to substrate distance of \~22 cm. Ar:N\textsubscript{2} ratio was 3:1 and resulted in a metallic mode. Targets were pre-sputtered under deposition conditions with gun shutters closed for 6 minutes prior to deposition of a film. A substrate temperature of 550 °C was used. Compositions of the films used for HAXPES measurements were estimated based upon deposition rates and Scanning Electron Microscopy (SEM) Energy Dispersive Spectroscopy (EDS) measurements.

\section{Computational Methods}

\subsection{Piezoelectric Coefficient}
We use Vienna Ab-initio Simulation Package (VASP 5.4.4) to perform calculations based on density functional theory (DFT).\cite{kresse1996} Plane-wave basis set with cutoff kinetic energy of 340 eV is used. The electron-ion interactions are described by projector-augmented wave (PAW) method.\cite{PAW} Specifically, we used the following VASP distributed pseudopotentials: Al 04Jan2001, N\textunderscore s 07Sep2000, and Hf\textunderscore pv 06Sep2000. The Perdew-Burke-Ernzerhof (PBE) exchange correlation functional\cite{perdew1996} was used and an on-site Hubbard potential $U$ = 3.0 eV was applied to the Hf $d$ orbitals. 
For each Hf composition, we calculate the piezoelectric coefficient and elastic constant tensors based on the fully relaxed structures, including the degree of freedom of cell shape and size, and ionic positions, and report the statistics out of four different structures (average and standard deviation). The self-consistent calculations of response to finite electric field are performed using the finite difference approach. Electric field vector (E$_x$,E$_y$,E$_z$) = (0.01, 0.01, 0.01) eV/\AA \ is used with fourth order finite difference stencil. A $\mathrm{\Gamma}$-centered Monkhorst-Pack $4\times4\times2$ \textbf{\textit{k}} grid and stricter self-consistency cutoff energy of 10$^7$ eV are used to ensure converged results.
For ionic contributions to piezoelectric coefficient and elastic constant tensors, ionic displacement of 0.01 \AA \ and the central difference approach is used. VASP calculate the piezoelectric strain coefficients (e$_{ij}$) directly and were converted to the piezoelectric stress coefficients (d$_{ij}$) using elastic constant tensor (C$_{ij}$).



\subsection{Defect Calculation}
All the defect calculations were performed using the VASP6 code\cite{kresse1996} with the PAW method.\cite{PAW} We adopt the hybrid functional (HSE06) to accurately calculate the defect formation energy of AlN.\cite{Krukau_JCP_2006} The cutoff kinetic energy is set to 520 eV. The defect formation energies ($\Delta E_{D,q}$) were calculated by using following equation,
 \begin{equation}
     \Delta E_{D,q} = E_{D,q}-E_{\rm{bulk}}-\sum_in_i\mu_i+q(\epsilon_{\rm{VBM}}+\mu_e)+E_{\rm{corr}}
 \end{equation}
 where $E_{D,q}$ and $E_{\rm{bulk}}$ are total energies of the supercell with a point defect $D$ in charge state $q$ and perfect lattice, respectively. These energies are calculated with a 3×3×2 supercell using a 2×2×2 $k$-point grid. The point defects considered in this work are the Al vacancy ($V_{\rm{Al}}$) in $q = -1, 0, 1, 2, 3, 4$, the N vacancy ($V_{\rm{N}}$) in $q = -4, -3, -2, -1, 0, 1$, and the substitutional defect of Hf-Al ($\rm{Hf_{Al}}$) in $q = -1, 0, 1, 2, 3$. In the equation above, $n_i$ represents the number of element $i$ that are added to the system. In addition, $\mu_i$ denotes the chemical potential of element $i$. To determine the region of allowed chemical potential, we calculated a chemical potential diagram of Al-Hf-N ternary system at 0K. The calculated values of chemical potential are listed in Table \ref{tab:chemical}. The energy level of the valence band maximum is represented by $\epsilon_{\rm{VBM}}$ and $\mu_e$ is the position of the Fermi energy referenced to $\epsilon_{\rm{VBM}}$. Finally, $E_{\rm{corr}}$ is the image-charge correction term proposed by Kumagai and Oba.\cite{kumagai_PRB_2014} The calculation of $\Delta E_{D,q}$ were performed by using pydefect software.\cite{kumagai_PRM_2021}
 
 The concentration of defect $D^q$ is given by the following equation,
  \begin{equation}
     \lbrack D^q\rbrack =N_{\rm{site}}\exp(-\frac{\Delta E_{D,q}}{k_{\rm{B}}T})
 \end{equation}
 where $N_{\rm{site}}$ denotes the number of sites in which the defect can be incorporated, $k_{\rm{B}}$ is the Boltzmann constant, and $T$ indicates the temperature. Solving this equation in combination with the charge neutrality condition, using the Fermi energy ($\mu_{\rm{e}}$) as a variable, gives the equilibrium Fermi energy and defect concentration. The carrier concentration was obtained by multiplying the density of states of AlN with the Fermi-Dirac distribution function. The computation of defect/carrier concentration was performed by using py-sc-fermi software.\cite{buckeridge_CPC_2019}

\renewcommand{\thetable}{S\arabic{table}}
\begin{table}[h]
 \centering
 \caption{Values of the chemical potentials and the competing phases with AlN in a ternary Al-Hf-N system. Condition A and B corresponds to Al-rich and N-rich conditions, respectively.}
 \label{tab:chemical}
 \begin{tabular}{ccccc} \hline
 Condition & $\Delta\mu_{\rm{Al}}\,(\rm{eV})$ & $\Delta\mu_{\rm{Hf}}\,(\rm{eV})$ & $\Delta\mu_{\rm{N}}\,(\rm{eV})$ & Competing phase \\ \hline
 A & $0$ & $-1.54$ & $-3.13$ & $\rm{Al, HfAl_2}$ \\
 B & $-3.13$ & $-4.26$ & $0$ & $\rm{N_2, Hf_3N_4}$ \\ \hline
 \end{tabular}
\end{table}

\begin{figure*}[h]
\includegraphics[scale=0.5]{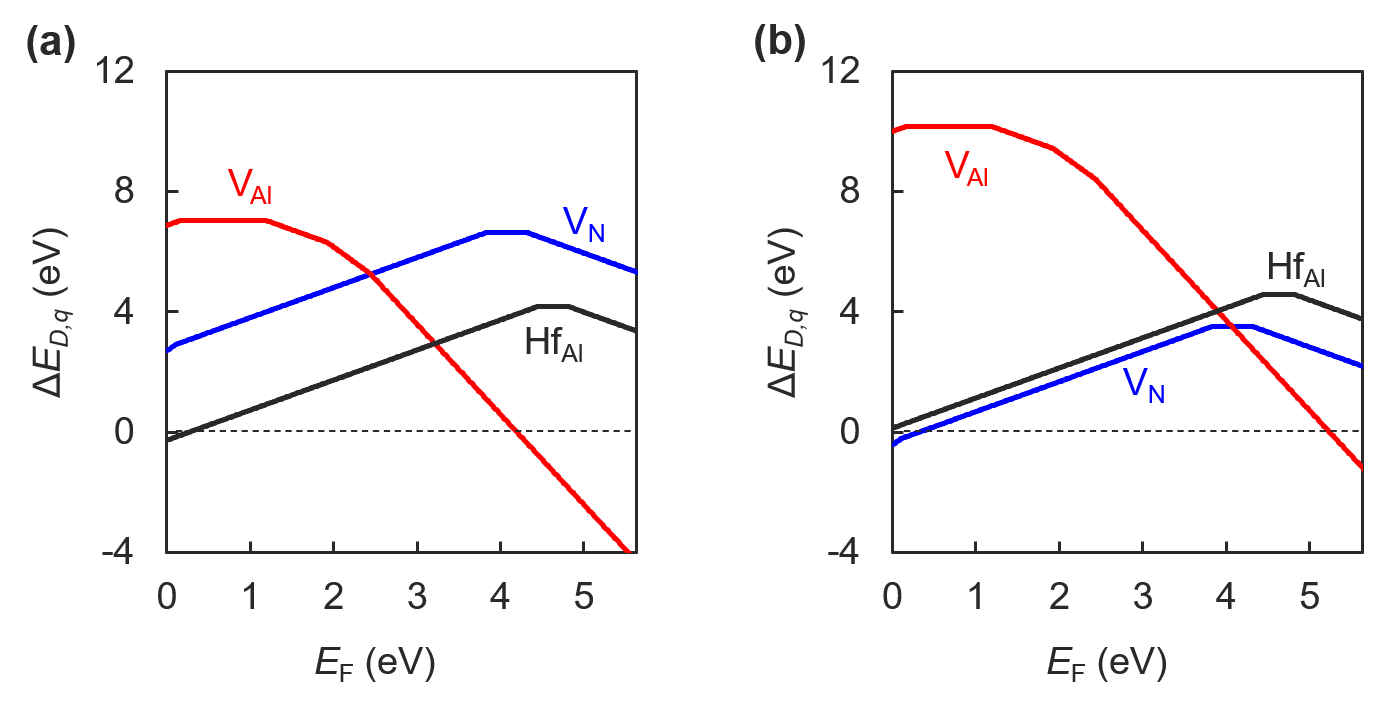}
\caption{\label{fig:SM_dft_defect} Defect formation energy ($\Delta E_{D,q}$) as a function of fermi energy ($E_{\rm{F}}$) in AlN under (a) N-rich and (b) Al-rich growth conditions.}
\end{figure*}

\bibliographystyle{apsrev4-1}
\bibliography{AlHfN_APL}